\documentclass[prl,twocolumn,twoside,showpacs,floatfix]{revtex4}
\usepackage{epsf,times}
\begin{document}
\draft
\title{Emerging behavior in electronic bidding}
\author{ I. Yang$^1$, H. Jeong$^2$, B. Kahng$^1$, and A.-L. Barab\'asi$^3$ \\}
\affiliation{
\mbox{$^1$ School of Physics and Center for Theoretical Physics,
Seoul National University, Seoul 151-747, Korea} \\
\mbox{$^2$ Department of Physics, Korea Advanced Institute of Science and 
Technology, Taejon, 305-701, Korea} \\
\mbox{$^3$ Department of Physics, University of Notre Dame, Notre Dame, IN 
46556, USA} }
\date{\today}
\begin{abstract}
We characterize the statistical properties of a large number of agents 
on two major online auction sites. The measurements indicate that 
the total number of 
bids placed in a single category and the number of distinct auctions 
frequented by a given agent follow power-law distributions, implying
that a few agents are responsible for a significant fraction of the 
total bidding activity on the online market. We find that these 
agents exert an un-proportional influence on the final price of 
the auctioned items. 
This domination of online 
auctions by an unusually active minority may be a generic feature 
of all online mercantile processes. 
\end{abstract}
\pacs{89.75.-k, 89.75.Da.}
\maketitle
Electronic commerce (E-commerce) is any type of business or commercial 
transaction that involves information transfer across the 
Internet. Over the past five years, E-commerce has expanded 
rapidly, taking the advantage of faster, cheaper and more convenient 
transactions over traditional ways. 
A synergetic combination of the Internet supported instantaneous 
interactions and traditional auction mechanisms, online auctions 
represent a rapidly expanding segment of E-commerce. 
Indeed, with the advent of the Internet most limitations of 
traditional auctions, such as 
geographical and time constraints, have virtually disappeared, making a 
significant fraction of the population potential auction participants 
\cite{1,2}. For example eBay, the largest consumer-to-consumer auction site, 
boosts over 40 million registered consumers, and has grown in revenue 
over 100,000 percent in the past five years. With the rapidly increasing 
number of agents the role of individuals diminishes and self-organizing 
processes increasingly dominate the market's behavior~\cite{3,4}. 
On the other hand, recently the self-organizing features of 
complex systems have attracted the attention of the statistical 
physics community because they contain diverse cooperations 
among numerous components of a system, resulting in patterns and 
behavior which are more than the sum of the individual action 
of the components. 
While many systematic studies have been carried out to understand 
such emerging patterns in various systems, little attention has 
been paid to electronic auctions. 
In this Letter, we collect auction data and 
show that the bidding of hundreds of thousands of agents 
leads to unexpected emerging behavior, impacting on everything 
from the bidding patterns of the participating agents to the final 
price of the auctioned item. We found that the total number of 
bids placed in a single category by a given agent follows 
a power-law distribution. The power-law behavior is 
rooted in the finding that an agent that makes frequent bids 
up to a certain moment is more likely to bid 
in the next time interval. Moreover, we find that 
the number of distinct items frequented by a given agent also 
follows a power-law distribution. 
The power-law behavior implies that a few powerful 
agents bid more frequently and on more distinct items than 
others. We will show that such powerful agents exert strong influence 
on the final prices in distinct auctions.         

To be specific, we collected auction data from two different 
sources. First, we downloaded 
all auctions closing on a single day on eBay, including 264,073 auctioned 
items, grouped by the auction site in 194 subcategories. The dataset allowed 
us to identify 384,058 distinct agents via their unique user ID. To verify 
the validity of our findings in different markets and time spans, we 
collected data over a one year period from eBay's Korean partner, 
auction.co.kr, involving 215,852 agents that bid on 287,018 items in 62 
subcategories. 

\begin{figure}
\centerline{\epsfxsize=7.3cm \epsfbox{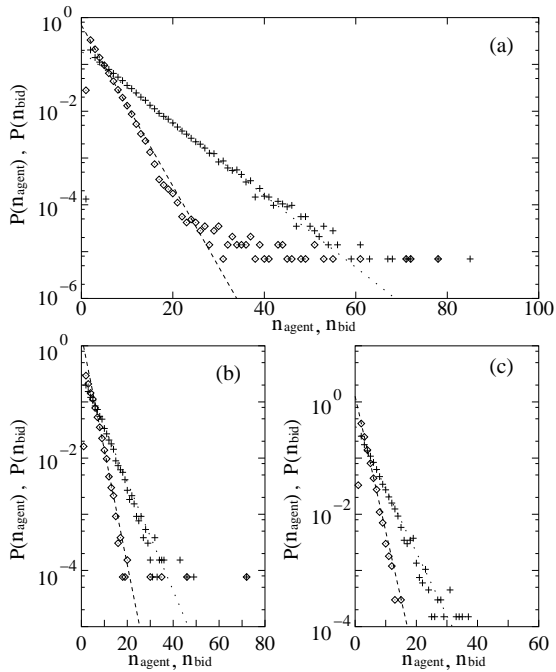}} 
\caption{
Bid and agent distribution on eBay. (a) Distribution of number of agents 
($n_{\rm agent}$,($\diamond$)) simultaneously bidding on a certain item and 
number of bids ($n_{\rm bids}$,(+)) received for an item, obtained by 
considering all items contained in the 194 categories on individual bids 
that were collected from auctions ending on July 5, 2001 on eBay. 
(b) and (c) Agent and bid distribution in the largest (b) and the 
second largest (c) category on eBay. The largest category by the 
number of auctioned items contains 21,461 items related to sport 
trading cards while the second largest category includes 13,610 items 
related to printed and recorded music. The straight lines correspond 
to exponential fits and the symbols are the same as in (a).
}
\label{fig1}
\end{figure}

\begin{figure}
\centerline{\epsfxsize=7.3cm \epsfbox{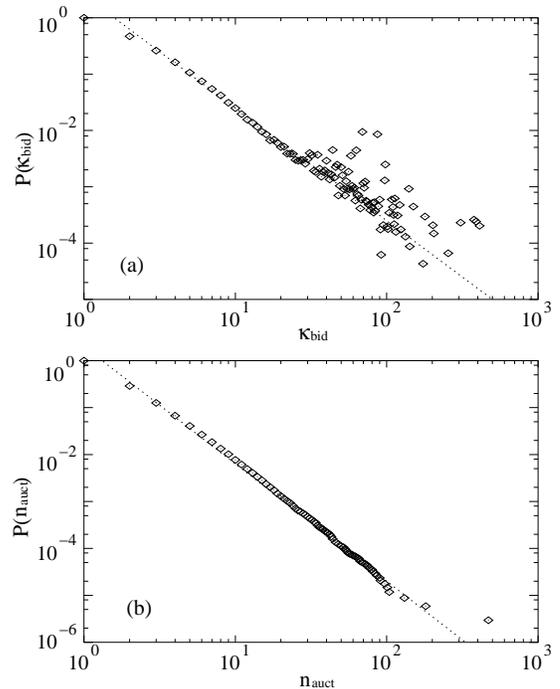}} 
\caption{
Frequency of bids placed by individual agents. (a) Cumulative distribution 
of total number of bids, $\kappa_{\rm bid}$, placed by a given agent in 
auctions in the same category. For each of the 194 categories we separately 
determined the cumulative distributions and averaged the obtained curves. 
(b) Cumulative distribution of the number of distinct auctions, $n_{\rm auct}$, 
frequented by a given agent. The dotted line on (a) has slope -2 while 
on (b) has slope -2.5, indicating that the corresponding probability 
distribution follows $P(\kappa_{\rm bid}) \sim \kappa_{\rm bid}^{-3}$ 
and $P(n_{\rm auct}) \sim n_{\rm auct}^{-3.5}$, respectively.
} 
\label{fig2}
\end{figure}

\begin{figure}
\centerline{\epsfxsize=7.3cm \epsfbox{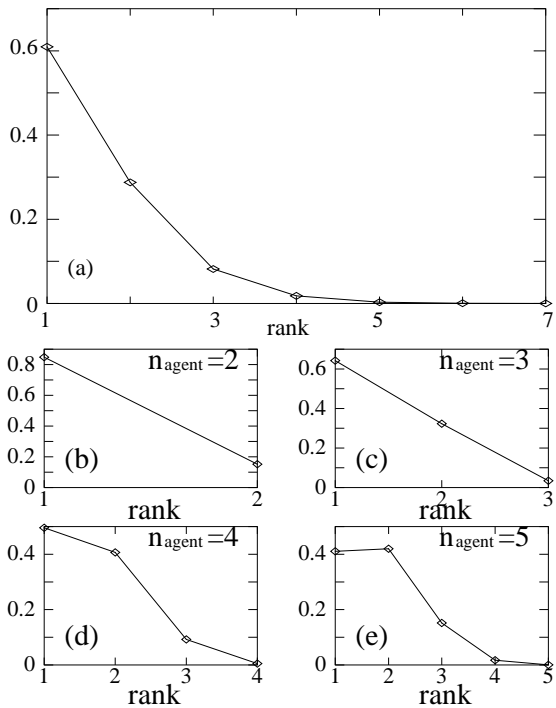}} 
\caption{
Frequent bidders more likely to win an auction. (a) The probability 
that an auction is won by agents with given activity rank. Using all 
completed auctions we calculated how many times the most, the second, 
or the n-th frequent bidder wins the auction. (b) The probability that 
the most frequent bidder wins the auction of two participants. (c)-(e): 
the probability that an agent wins an auction of n [(c) n=3; (d) n=4; 
(e) n=5] participants. (a) is based on 143,325 auctions, while (b)-(e) 
are based on 47,610, 30,205, 20,017, and 13,762 auctions, respectively.
} 
\label{fig3}
\end{figure}

\begin{figure}
\centerline{\epsfxsize=7.3cm \epsfbox{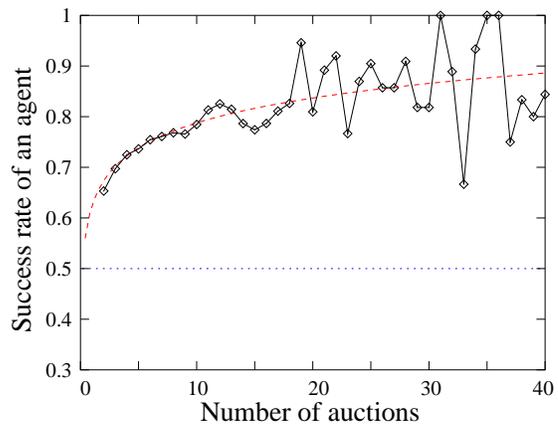}} 
\caption{
The dependence of an agent¡¯s success rate on the number of auctions 
the agent participates. For each product subcategory (containing highly 
similar items) we calculated $P_i^{\rm win}$, the average of the winning 
prices for items won by agent $i$. For the same agent we also calculated 
$P_i^{\rm lost}$, the average over the winning price over items in which 
agent $i$ participated but lost. A successful agent can get a lower price 
for the items it won than other agents bidding on similar items on parallel 
auctions, i.e. for a successful agent $P_i^{\rm win}$ $<$ $P_i^{\rm lost}$. 
We find that the success rate of an agent, measured as the function of 
auctions won at a lower than average price (i.e. the fraction of agents 
for which $P_i^{\rm win}$ $<$ $P_i^{\rm lost}$) increases with the number 
of auctions these agents participate in. A horizontal dotted line corresponds 
to the case when there is no correlation between the frequency of bidding 
and the chances of getting a better price. A numerical fitting indicates 
that the success rate increases logarithmically (dashed line). 
} 
\label{fig4}
\end{figure}

\begin{figure}
\centerline{\epsfxsize=7.3cm \epsfbox{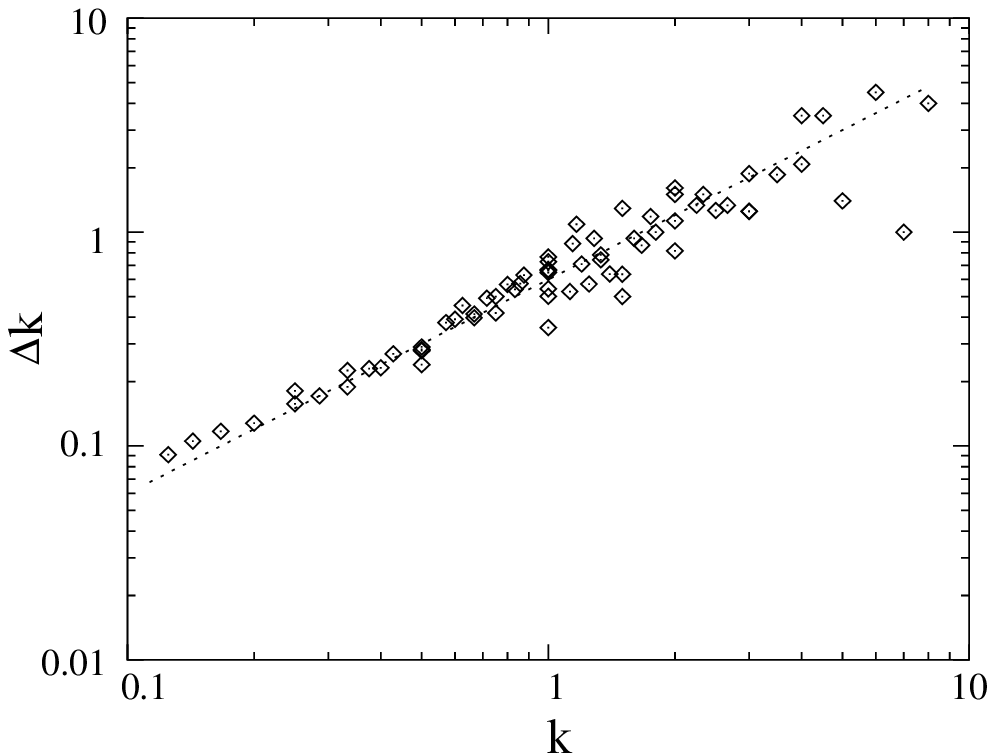}} 
\caption{
The origin of the observed power law is in preferential attachment. 
The figure shows the change in the number of bids placed by an agent $i$ 
that previously placed $k$ bids, averaged over all items and all times. 
The linear behavior in the log-log plot indicates that the more bids 
an agent places up to a given moment, the more likely it is that it 
will place another bid in the next time interval. Such preferential 
bidding is known to lead to a power law bidding distribution [7,9].
The dotted line with slope 1 is shown for visual reference. 
} 
\label{fig5}
\end{figure}

In a typical online auction a seller places the item's description on the 
auction site and sets the starting and the closing time for the auction. 
Agents (bidders) submit bids for the item. Each new bid has to exceed the 
last available bid by a preset increment. Agents can bid manually, placing a 
fixed bid, or on some auction sites (such as eBay but not on their Korean 
partner) they can take advantage of proxy bidding. In proxy bidding an agent 
indicates to the auction house the maximum price he/she is willing to pay 
for the given item (proxy bid), which is not disclosed to other bidders. 
Each time a bidder increases the bid price, the auction house makes 
automatic bids for the agent with an active proxy bid, outbidding the last 
bid with a fixed increment, until the proxy price is reached. In online 
consumer-to-consumer auctions the agent with the highest bid wins and pays 
the amount of that bid; all other participants pay nothing. 

Most online auction sites keep a detailed, publicly available record of all 
bids and identify the bidding agents via an unique login name. It is this 
transparency of the bidding history that allows us to characterize in 
quantitative terms the auction process. Each completed auction can be 
characterized by two quantities: the number of distinct agents bidding on 
the same item ($n_{\rm agent}$) and the total number of recorded bids for the 
item ($n_{\rm bids}$), where $n_{\rm bids} \ge n_{\rm agent}$, as each 
agent can place multiple bids. In Fig.~1 we show the distribution of 
$n_{\rm agent}$ and $n_{\rm bids}$ over all auctions recorded on eBay, 
finding that they both follow $P(n)\sim \exp(-n/n_{0})$, where $n_0=5.6$ 
for $n_{\rm bids}$ and $n_0=2.5$ for $n_{\rm agent}$. 
We obtained similar results for the Korean market, with $n_0=10.8$ 
for $n_{\rm bids}$ and $n_0=7.4$ for $n_{\rm agent}$. 
This simple exponential form is unexpected, as one expects that the bidding 
distribution is the result of many independent events, and therefore follows 
a Gaussian, peaked around the average number of bids and decreasing as 
$\sim \exp(-an^2)$ with a constant $a$. 
The deviation from a Gaussian distribution 
could come from the fact that Fig.~1a collapses data from different 
categories, displaying different bidding patterns. In Fig.~1b and 
c we show the distribution in two subcategories 
(sports trading cards and printed, 
recorded music), finding that they follow the same functional form as the 
aggregated data. Therefore, the exponential form for the activity 
distribution appears to be a general feature of all auctions, indicating 
that the majority of auctions have only a few bidders and auctions with a 
large number of bids or participating agents are exponentially rare.

To characterize the activity of individual agents we determined the number 
of bids placed by each agent on each auction. As agents place simultaneous 
bids on items that closely resemble each other, we denote by 
$\kappa_{\rm bid}$ the total number of bids placed by the same agent 
in auctions in the same category. For example, if several similar computers 
are sold on separate auctions, agents looking for a computer often bid 
simultaneously for several or all of them. We find that the distribution of 
$\kappa_{\rm bid}$ follows a power law 
\begin{equation}
P(\kappa_{\rm bid})\sim \kappa_{\rm bid}^{-\gamma},
\end{equation} 
where $\gamma=3$ (Fig.~2a) on both eBay and the Korean auction. 
A similar power law characterizes the distribution of the 
number of different auctions, $n_{\rm auct}$, frequented by individual agents, 
finding that 
\begin{equation}
P(n_{\rm auct}) \sim n_{\rm auct}^{-\beta},
\end{equation}
where $\beta=3.5$ (Fig.~2b). 
The power-law distribution shown in Fig.~2a implies that 
while most agents place only a small number of bids, a few agents bid very 
frequently, placing several hundred bids on the same day. Similarly, Fig.~2b 
indicates that while most agents participate in a few auctions only, a few 
agents bid very widely, some placing simultaneous bids on over a hundred 
distinct items on the same day. Therefore, unknown to most participants, 
the auction process is dominated by a small number of highly active agents, 
or \textit{power-agents,} that pursue a very aggressive bidding pattern, 
placing simultaneously a large number of bids on a wide range of items. 
These power-agents are responsible for the power-law tail of the 
distribution shown in Fig.~2. Our measurements indicate that there 
is a strong correlation between the number of bids placed by an agent 
on an item and the number of items the same agent bids for, indicating 
that power agents simultaneously bid frequently and widely. 

Agents with an aggressive bidding pattern significantly alter the nature of 
the bidding process, potentially distorting the chance of a typical agent to 
win an auction. To inspect the effect of the bidding pattern on the success 
rate of a given agent, we determined the fraction of auctions won by the 
most, the second, or the $k$-th most frequently bidding agents. 
We find that in 61{\%} of all auctions the winner is the agent that 
makes the most bids, and in 29{\%} of all auctions the second most 
frequently bidding agent wins the auction (Fig.~3a). Less than 
0.3{\%} of the auctions are won by agents whose 
activity ranks fifth or higher. As most auctions have only a few 
participating agents (Fig.~1a), it is useful to re-examine the winning 
patterns in auctions with the same number of agents. We find that if only 
two agents participate in an auction, and each place multiple bids, in 
85{\%} of the cases the agent with more bids wins the auction (Fig.~3b). 
The situation is similar for three agents as well (Fig.~3c): in 64{\%} of the 
cases the more active agent wins, followed by the second most active, which 
wins in 32{\%} of the cases. In auctions with larger number of participants 
(Fig.~3d-e) we observe a similar pattern. These results indicate that 
frequently bidding agents play a key role in setting the final price of most 
auctioned items: in the vast majority of the cases the agents who place the 
largest number of bids are the winners of the auction process. This finding 
indicates that despite the widespread practice of sniping among experienced 
users \cite{5} when bidders place bids only in the last 60 seconds of the 
auction hoping to win the auctioned item, on average frequent bidders are 
more successful. 

As the power law distribution (Fig.~2b) indicates that some agents bid 
rather widely, the question is, does such wide bidding result in economic 
advantage for power bidders? Our results indicate that power agents not only 
are the frequent winners of the auctions in which they participate, but they 
also pay less than other agents on similar items. Indeed, in Fig.~4 we show 
the fraction of times the most frequently bidding agent pays less for an 
item than other agents that win auction of items in the same category. We 
find that the more auctions an agent participates, the larger is its chance 
to pay less for the same item than the less widely bidding agents.

The observed power law distribution is rooted in the dynamics of the bidding 
process. Indeed, two processes contribute to the final number of bids placed 
on a given item: new agents entering the bidding process and agents that 
already placed a bid increase their bid. We find that this pattern is 
governed by a process often referred to as preferential attachment, similar 
to those responsible for the emergence of scaling in complex networks~\cite{6}:
more bids an agent places on a given item up to a certain moment, 
more likely is that he/she will place an another bid in the future (Fig.~5). 
The linearity of the observed relationship is known to lead to power law 
distributions, as demonstrated by studies in both economics~\cite{7,8} 
and complex networks~\cite{9,10,11}.

While power laws have been often observed in economic contexts, ranging from 
city \cite{8} and company size distribution \cite{12,13} to Pareto's 
observation of wide income distributions \cite{14} and time series 
analysis \cite{5}, they are rather unexpected during the frequency of 
bidding of individual users. In order to develop an analytical framework 
to capture the dynamics of the bidding process current 
auction models inevitably make use of equilibrium concept \cite{15,16}. 
Often this requires the assumption that the number of agents is 
fixed \cite{15} which, while leads to analytically tractable models, 
is not realistic in the context of Internet auctions. 
Indeed, the power laws observed here are the result of 
the auction's fundamental openness and non-equilibrium nature. In the past 
few years the observation of such non-equilibrium features in economic 
phenomena has led to an increased interest among physicists and 
mathematicians in the self-organizing processes governing economic systems 
\cite{3,4,12,17,18}. Our finding, that similar non-equilibrium processes 
govern the behavior of online auctions places these mercantile processes 
in the realm of agent driven self-organization. 

In conclusion, we have collected online auction data and analyzed 
the statistical properties of emerging patterns created by a large 
number of agents. We found that the total number of bids placed 
in a single category and the number of distinct auctions frequented 
by a given agent follow power-law distributions. Such power-law 
behaviors imply that the online auction system is driven 
by self-organized processes, involving all agents participated 
in a given auction activity. We also uncovered the empirical fact 
that the more bids an agent places up to a given moment, more 
likely it is that it will place another bid in the next time interval, 
which plays an important role in generating the power-law behavior in 
the bidding frequency distribution by a given agent.\\  

This work is supported by the KOSEF Grant No. R14-2002-059-01000-0 in 
the ABRL program.

\end{document}